\documentclass[11pt]{article}
\usepackage{graphicx}
\begin{document}
\title{\bf  On a complementary scale of crystal-field strength}
\author{\bf J. Mulak$^{1}$ and M. Mulak$^{2}$}
\date{{\it  $^{1}$ W Trzebiatowski Institute of Low Temperature
            and Structure Research,\\
            Polish Academy of Sciences, 50--950, PO Box 1410,
            Wroclaw, Poland\\
            $^{2}$ Institute of Physics,
            Wroclaw University of Technology,\\
            Wyb. Wyspianskiego 27,
            50--370 Wroclaw, Poland}}
\maketitle
\begin{abstract}
A new measure of the crystal-field strength, complementary to the conventional one, is defined. It is based on
the rotational invariants $\left|B_{k0}\right|_{\rm av}$ or $\left|\sum_{k}B_{k0}\right|_{\rm av}$, $k=2,4,6$,
of the crystal-(ligand)-field (CF) Hamiltonian ${\cal H}_{\rm CF}$ parametrizations, i.e. on the axial CF
parameters modules averaged over all reference frame orientations. They turn out to be equal to $\left|{\cal
H}_{\rm CF}^{(k)}\right|_{\rm av}$ and $\left|{\cal H}_{\rm CF}\right|_{\rm av}$, respectively. While the
traditional measure is established on the parametrization modules or on the second moment of the CF energy
levels, the introduced scale employs rather the first moment of the energy modules and has better resolving
power. The new scale is able to differentiate the strength of various iso-modular parametrizations according to
the classes of rotationally equivalent parametrizations. Using both the compatible CF strength measures one may
draw more accurate conclusions about the Stark levels arrays and particularly their total splitting magnitudes.
\end{abstract}

\newpage

\section*{1. Introduction}
Solid state experimentalists, especially spectroscopists, still need a reliable scale quantitatively
characterizing the effect of crystal-field interaction, i.e. defining the so-called crystal-field strength. Such
a parameter could directly verify and compare various parametrizations of the crystal-field Hamiltonian ${\cal
H}_{\rm CF}$, which may come from different fittings experimental data when the orientations of reference frames
associated with these parametrizations are unknown in the majority of cases.

Although such a conventional scale for measuring the strength of the crystal-field has been already introduced
over twenty years ago [1,2], in some cases it seems to be insufficiently precise. It employs the basic
rotational invariants of the ${\cal H}_{\rm CF}$, i.e. the modules of its $2^{k}$-pole components ${\cal H}_{\rm
CF}^{(k)}$, defined as $M_{k}=\left(\sum_{q}|B_{kq}|^{2}\right)^{1/2}$, as well as uses the global ${\cal
H}_{\rm CF}$ modulus $M=\left(\sum_{k}\sum_{q}|B_{kq}|^{2}\right)^{1/2}$. In the first case the partial
crystal-field strength is defined as $S_{k}\!=\!\left(\frac{1}{2k+1}\right)^{1/2}M_{k}$, while in the second
case the global crystal-field strength is given by $S=\left(\sum_{k}S_{k}^{2}\right)^{1/2}$. Throughout the
paper the tensor (Wybourne) notation for the crystal-field Hamiltonian and the crystal-field parameters (CFPs),
${\cal H}_{\rm CF}=\sum_{k}\sum_{q}B_{kq}C_{q}^{(k)}$, is consistently used [3]. The summations over $k$ and $q$
indices run, in each individual case, over strictly specified values according to the kind of central ion and
its point symmetry.

Both the parameters $S_{k}$ and $S$ themselves are not a direct measure of the real magnitude of the initial
state splitting, since the crystal-field effect depends also on the properties of an object (a paramagnetic ion)
upon which the ${\cal H}_{\rm CF}$ acts. Namely, the response of the system to the ${\cal H}_{\rm CF}$
perturbation reflects the symmetry of the electron density distribution of the central ion open-shell. For
instance, an $S$-type ion like Gd$^{3+}$ feels no crystal field (in the first order of perturbation) no matter
how strong is the surrounding field.

The effect of splitting can be most simply expressed by the so-called second moments $\sigma_{k}^{2}$ or
$\sigma^{2}$ of the CF sublevels within the initial state upon switching on the ${\cal H}^{(k)}_{\rm CF}$ (or
${\cal H}_{\rm CF}$) perturbation. In fact, the second moment is easily represented by the scalar crystal-field
strength parameters, either $S_{k}$ or $S$ (section 2). However, although the effective ${\cal H}^{(k)}_{\rm
CF}$ multipoles (for $k=2,4,6$) contribute to the energy of individual Stark levels independently (as an
algebraic sum), the simple linear relations between $\sigma_{k}^{2}$ (or $\sigma^{2}$), and $S_{k}^{2}$ (or
$S^{2}$) are always fulfilled. As it is proved these relations strongly confine both the maximal $(\Delta {\cal
E}_{\rm max})$ and minimal $(\Delta {\cal E}_{\rm min})$ nominally allowed splittings of the initial state
(section 3). Moreover the actual crystal field splittings $\Delta E$ can be additionally restricted (section 5).
Naturally, all the iso-modular ${\cal H}_{\rm CF}$ parametrizations correspond to the same crystal-field
strengths $S_{k}$ and $S$. However, apart from the modules $M_{k}$ and $M$, there exist also other rotational
invariants of the ${\cal H}^{(k)}_{\rm CF}$ or ${\cal H}_{\rm CF}$ which distinguish the whole classes of the
rotationally equivalent ${\cal H}_{\rm CF}$ parameterizations, in other words the parameterizations referring to
the same real crystal-field potential, but expressed in variously oriented reference frame. Interestingly, the
new invariants turn out to be the average values of the axial parameter modulus $|B_{k0}|_{\rm av}$, $k=2,4,6$,
in the case of ${\cal H}^{(k)}_{\rm CF}$, or $|\sum_{k}B_{k0}|_{\rm av}$ for the global ${\cal H}_{\rm CF}$
obtained after the averaging over all orientations of the reference frame, i.e. over the solid angle $4\pi$. As
it is shown in the paper the average value of the axial parameter modulus or the average of the modulus of their
sum are just equal to $|{\cal H}^{(k)}_{\rm CF}|_{\rm av}$ and $|{\cal H}_{\rm CF}|_{\rm av}$, respectively
(section 4).

The new scale of the crystal-field strength based on the above invariants is in principle consistent with the
conventional one but it reveals more resolving power. Applying the new measure to the iso-modular
parametrizations may lead to different strength parameters what is exemplified below for several cases (section
5). The introduced more subtle strength gradation established rather on the first moment of the sublevel energy
modules gives, comparing to the second moment, additional information about the Stark levels array for various
iso-modular ${\cal H}_{\rm CF}$s, including the magnitude of the total splitting gap of the states. In this
paper we confine ourselves to the pure model states of the zero-order approximation with a well defined angular
momentum quantum number and the corresponding degeneration. These could be for instance Russell-Saunders coupled
states $|\alpha L S J \rangle$ coming from the $^{2S+1}$L terms, where $\alpha$ stands for the remaining quantum
numbers needed for their complete determination. Such states have a well defined quantum number $J$ and the
degeneration $2J+1$.  The derivation of the analogical expressions including $J$-mixing effects [4] or a
transformation to other functional bases of the zero-order approximation can be accomplished by using standard
angular momentum re-coupling techniques [4-8]. In section 5 we analyse by way of example the crystal-field
splitting of $p^{1}$, $d^{1}$ and $f^{1}$ one-electron configurations and a typical complex state $^{3}H_{4}$
for various iso-modular ${\cal H}_{\rm CF}^{(k)}$, $k=2,4,6$. In the first three cases we avoid complex states
re-coupling procedure which is a side issue to the problem under consideration. Since we study the
differentiation of the effects due to various iso-modular Hamiltonians ${\cal H}_{\rm CF}^{(k)}$, all CFPs
values along with the Stark levels energies are given in $M_{k}$ units.

\subsection*{2. Conventional definition of the crystal-field strength parameter}
The comparison and scaling of the crystal-field impact can be based upon the two types of scalar quantities,
$M_{k}$ and/or $M$, since both of them are rotationally invariant. A scalar crystal-field strength parameter of
this kind was given firstly by Auzel and Malta [1,2] as (in original notation):
\begin{eqnarray*}
  N_{v} &=& \left[\sum\limits_{k,q}|B_{q}^{k}|^2\left(\frac{2\pi}{2k+1}\right)\right]^{1/2},
\end{eqnarray*}
which is nothing more but $M$ in the space spanned by spherical harmonics $Y_{q}^{k}$. In other words, $N_{v}$
is a norm representing a distance in the space. Currently there are two definitions widely used in the
literature [9-12]:
\begin{eqnarray}
  S_{k} &=& \left(\frac{1}{2k+1}\sum\limits_{q}|B_{kq}|^2\right)^{1/2}=
  \left\{\frac{1}{2k+1}\left[B_{k0}^{2}+2\sum\limits_{q>0}({\rm Re}B_{kq})^2+({\rm
  Im}B_{kq})^2\right]\right\}^{1/2},
\end{eqnarray}
for $k=2,4$ and $6$ in the case of $2^{k}$-pole ${\cal H}_{\rm CF}$ component and
\begin{eqnarray}
  S &=& \left(S_{2}^{2}+S_{4}^{2}+S_{6}^{2}\right)^{1/2} \qquad {\rm or} \qquad
  S=\left[\frac{1}{3}\left(S_{2}^{2}+S_{4}^{2}+S_{6}^{2}\right)\right]^{1/2},
\end{eqnarray}
for the global ${\cal H}_{\rm CF}=\sum_{k}{\cal H}_{\rm CF}^{(k)}$ [4]. A word of caution seems to be worthy at
this point. Namely, the values of $S_{k}$ or $S$ can differ according to the type of the ${\cal H}_{\rm CF}$
parametrization (operators) applied. They can be compared with each other only after proper recalculation. Since
both these quantities are independent of the assumed axis system they allow to check whether the original CFP
data sets and the transformed ones are compatible. The strengths $S_{k}$ or $S$ enable also a broad comparison
of CFP data sets when the axis systems have not been explicitly defined, and undoubtedly they play a central
role in the CF theory. What is also important and useful they are linked to the second moment of the Stark
levels within a particular initial state $|\alpha S L J\rangle$ [4,13].

The second moment of the sublevels $|n\rangle$ within $|\alpha S L J\rangle$ state upon introduction of a ${\cal
H}_{\rm CF}$ perturbation is defined by
\begin{eqnarray}
  \sigma^{2}(|\alpha S L J\rangle) &=& \frac{1}{2J+1}\sum\limits_{n}\left[E_{n}-\bar{E}(|\alpha S L J\rangle)\right]^{2},
\end{eqnarray}
where the center of gravity of the Stark levels belonging to the state $|\alpha S L J\rangle$ is given by
\begin{eqnarray*}
  \bar{E}(|\alpha S L J\rangle)&=& \frac{1}{2J+1}\sum\limits_{n}E_{n},
\end{eqnarray*}
and $E_{n}$ is the $|n\rangle$ sublevel energy. Since ${\cal H}_{\rm CF}$ is diagonal in the $|n\rangle$ basis
and the second order effect of ${\cal H}_{\rm CF}$ interaction is neglected [4]
\begin{eqnarray*}
  \sigma^{2}(|\alpha S L J\rangle) &=& \frac{1}{2J+1}Tr\left\{{\cal H}_{\rm CF}^{2}\right\}.
\end{eqnarray*}
Hence
\begin{eqnarray}
  \sigma^{2}(|\alpha S L J\rangle) &=& \frac{1}{2J+1}\sum\limits_{k}S_{k}^{2}
  \left(\langle\alpha SLJ||C^{(k)}||\alpha SLJ\rangle\right)^{2},
\end{eqnarray}
what implies from the orthogonality of 3-$j$ symbols [2,5,13]. The symbols $\langle\alpha SLJ||C^{(k)}||\alpha
SLJ\rangle$ are the double-bar or reduced matrix elements of the spherical tensor operators. According to
Wigner-Eckart theorem [14] they are independent of the reference frame orientation. Their origin and physical
meaning stem from the following relationships [5-8,15]:
\begin{eqnarray*}
  \langle\alpha SLJM_{J}|C_{q}^{(k)}|\alpha SL^{\prime}J^{\prime}M_{J}^{\prime}\rangle &=&
  (-1)^{J-M_{J}}\left(\begin{array}{ccc}
                  J & k & J^{\prime} \\
                 -M_{J} & q & M_{J}^{\prime} \\
                \end{array}\right)
\langle\alpha SLJM_{J}||C^{(k)}||\alpha SL^{\prime}J^{\prime}\rangle,
\end{eqnarray*}
where the reduced matrix element follows the 3-$j$ factor. Further use of tensor formalism yields
\begin{eqnarray*}
  \langle\alpha SLJ||C^{(k)}||\alpha SL^{\prime}J^{\prime}\rangle &=&
  (-1)^{S+L^{\prime}+J+k}
  \left[(2J+1)(2J^{\prime}+1)\right]^{1/2}\left\{\begin{array}{ccc}
                  J & J^{\prime} & k \\
                 L^{\prime} & L & S \\
                \end{array}\right\}
\langle\alpha SL||C^{(k)}||\alpha SL^{\prime}\rangle,
\end{eqnarray*}
where the double-reduced matrix element follows now the 6-$j$ symbol. We can also pass to the matrix elements of
the unit operators $U^{(k)}$ [5-8,15], i.e. normalized equivalents of $C^{(k)}$, since
\begin{eqnarray*}
   \langle\alpha SL||C^{(k)}||\alpha SL^{\prime}\rangle  &=&  \langle\alpha SL||U^{(k)}||\alpha SL^{\prime}\rangle
    \langle\l||C^{(k)}||l\rangle,
\end{eqnarray*}
and $l$ is the angular momentum quantum number of the open-shell electrons. The reduced matrix elements of the
$U^{(k)}$ operators have been compiled by Nielson and Koster [16], whereas the 3-$j$ and 6-$j$ symbols can be
found in the tables by Rotenberg et al [7].

The simple relation between the $\sigma_{k}^{2}$ and $S_{k}^{2}$ (Eq.4) can be also proved employing Vieta's
formulas for roots of the ${\cal H}_{\rm CF}^{(k)}$ matrix characteristic polynomial
\begin{eqnarray*}
  E^{n}+a_{1}E^{n-1}+a_{2}E^{n-2}+\ldots a_{n}&=& 0,
\end{eqnarray*}
which is here of order of $n=2J+1$. All its coefficients and roots must be real what follows obviously from the
${\cal H}_{\rm CF}$ hermiticity. Interestingly, some characteristics of the sublevels spectrum may be described
in terms of the elementary algebra. Firstly, as the energy center of gravity of the initial state is conserved,
i.e. $\left(\sum_{i=1}^{n}E_{i}=0\right)^2$, the $a_{1}$ coefficient standing at $E^{n-1}$ must vanish. Next,
since $0=\left(\sum_{i}^{n}E_{i}=0\right)^2=\sum_{i=1}^{n}E_{i}^{2}+ 2\sum_{i>j}E_{i}E_{j}$, the second moment,
i.e. the sum of the root squares (divided by $2J+1$) is equal to $\frac{-2a_{2}}{2J+1}$. It can be also shown
that
\begin{eqnarray*}
  -2a_{2} &=& \frac{1}{2k+1}\;M_{k}^{2}\;\langle J||C^{(k)}||J\rangle^{2},
\end{eqnarray*}
where the simplified notation for the reduced matrix element representing only the last quantum number has been
introduced. Hence, between $\sigma_{k}^{2}$ and $S_{k}^{2}$ a simple formula holds (Eq.4)
\begin{eqnarray}
  \sigma_{k}^{2} &=& \frac{1}{2J+1}\;S_{k}^{2}\;
  \langle J||C^{(k)}||J\rangle^{2}.
\end{eqnarray}
In other words, $\sigma_{k}$ is proportional to $S_{k}$. Finally, a free term of the characteristic polynomial
is given as $a_{n}=(-1)^{n}E_{1}E_{2}\ldots E_{n}$, what may be helpful analyzing the solutions. For instance,
if one root equals zero then a free term vanishes.

The problem becomes more complex for the global crystal-field strength $S$ (Eq.2), since then the components
$S_{k}^{2}$ contribute to the sum with their weights $\langle J||C^{(k)}||J\rangle^{2}$ (Eq.4). This is why
there is no straightforward proportionality between $\sigma^{2}$ and $S^{2}$ in this case. Nevertheless $\sigma$
is a positively defined quadratic form of $S_{k}$ and, in consequence, the inputs of particular ${\cal H}_{\rm
CF}$ $2^{k}$-poles into $\sigma^{2}$ cannot compensate themselves. The condition that $\sigma^{2}$ is constant
for various iso-modular ${\cal H}_{\rm CF}$ does not exclude, however, a possible differentiation of the CF
sublevels sequence and structure, as well as the initial state total splitting. In fact, $\sigma$ and $S$ could
be correlated similarly as $\sigma_{k}$ and $S_{k}$ in the previous case, only if the elements $\langle
J||C^{(k)}||J\rangle$ were equal for all $k$. Nevertheless, the second moment of the Stark levels within a
particular state $|\alpha S L J\rangle$ is simply given in terms of $S_{2}$, $S_{4}$ and $S_{6}$. Auzel and
Malta [2] made an attempt to average the $\sigma^{2}$ quadratic form by bringing down its respective ellipsoid
$\sum_{k}S_{k}^{2}\langle J||C^{(k)}||J\rangle^{2}$ in the $k$-space to a sphere of the same volume
$\sum_{k}S_{k}^{2}\left[\Pi_{k}\langle J||C^{(k)}||J\rangle^{2}\right]^{1/3}$ and having a radius equal to the
geometric mean of the three ellipsoid axes. In practice, unfortunately, this elegant approach does not always
lead to acceptable results. In the literature the overall effect of the crystal-field interaction is often
characterized by a quantitative comparison of the crystal-field strength [17-20]. Additionally, a systematic
correlation between the free ion parameters and the CF strength is observed, namely increase of the
crystal-field interaction results in the reduction of the free-ion parameters [17]. The CF strength increases in
the RE series with decreasing ionic radius of the RE$^{3+}$ host cation [19]. The physical meaning of the CF
strength scalar parameter is also supported by the fact that it rises with pressure applied to a sample [21,22].
The CF strength parameter has also been used to compare the root mean square error obtained for crystal fields
of different strength. However, its use in such a case is restricted only to comparisons of the identical site
symmetries [10,23]. Furthermore, within the approximation to the second order in the crystal-field, the shift in
the center of gravity of a particular $^{2S+1}L_{J}$ state due to $J$-mixing effects is a simple linear function
of the $S_{k}^{2}$ [4,17]. The concept of the $S_{k}^{2}$ or $S$ can be extended to define the quantities
$C_{k}$ and $C_{G}$ [13] as normalized "scalar products" of any two compared parametrizations. These quantities
represent the "angles" between the two considered parametrizations and are a convenient measure of the
closeness, i.e. the correlation of any two CFPs sets.

\subsection*{3. The correspondence of the Stark levels second moment of $|J\rangle$ state to its nominally
                allowed splittings}
The second moment of CF levels, $\sigma^{2}$, essentially limits a formally allowed range of the initial state
$|J\rangle$ total splittings $\Delta {\cal E}$ for different but iso-modular ${\cal H}_{\rm CF}$s. Such energy
splitting confinement differs for non-Kramers and Kramers ions what is specified in details below.

Let us firstly study the case of any integer $J$, i.e. non-Kramers ions. Having to keep a constant $\sigma^{2}$
the minimal hypothetical splitting $\Delta {\cal E}_{\rm min}$ of the $(2J+1)$-fold degenerate state takes place
when $J$ levels assume identical energy of $\frac{J+1}{2J+1}\Delta{\cal E}_{\rm min}$, and the remaining $J+1$
levels take the energy $\frac{-J}{2J+1}\Delta {\cal E}_{\rm min}$, or vice versa. Further this is referred as
Type I splitting. Then
\begin{eqnarray}
  \sigma^{2} &=& \frac{J(J+1)(\Delta {\cal E}_{\rm min})^{2}}{(2J+1)^{2}}, \qquad \qquad
  \Delta {\cal E}_{\rm min}=\sigma\frac{2J+1}{\sqrt{J(J+1)}}.
\end{eqnarray}
In turn, the maximal hypothetical splitting $\Delta {\cal E}_{\rm max}$ occurs for one level of $\Delta {\cal
E}_{\rm max}/2$ energy, one of $-\Delta {\cal E}_{\rm max}/2$, and the rest $(2J-1)$ levels with zero energy.
Further this is referred as Type II splitting. Then
\begin{eqnarray}
  \sigma^{2} &=& \frac{2\left(\Delta {\cal E}_{\rm max}/2\right)^2}{2J+1}, \qquad \qquad
  \Delta {\cal E}_{\rm max}=\sigma \sqrt{2(2J+1)},
\end{eqnarray}
and hence \begin{eqnarray}
 \frac{\Delta {\cal E}_{\rm max}}{\Delta {\cal E}_{\rm min}} &=& \sqrt{\frac{2J(J+1)}{2J+1}}.
\end{eqnarray}
Let us also consider, following Auzel and Malta [2], the case of the homogenous splitting $\Delta {\cal E}_{\rm
hom}$, when
\begin{eqnarray}
  \sigma^{2} &=& \frac{2(1+4+\ldots+J^{2})\left(\Delta {\cal E}_{\rm hom}/2J\right)^2}{2J+1} \qquad {\rm and} \qquad
  \Delta {\cal E}_{\rm hom}=2\sigma \sqrt{\frac{3J}{J+1}}.
\end{eqnarray}
Below it will be referred as Type III splitting. For example, if $J=4$ then $\Delta {\cal E}_{\rm
min}=\sigma\frac{9}{2\sqrt{5}}=2.01\sigma$, $\Delta {\cal E}_{\rm max}=\sigma3\sqrt{2}\sigma=4.24\sigma$,
$\Delta {\cal E}_{\rm hom}=4\sigma\sqrt{3/5}=3.10\sigma$, and finally the ratio $\frac{\Delta {\cal E}_{\rm
max}}{\Delta {\cal E}_{\rm min}} = 2.11$.

For $J=1$, $\Delta {\cal E}_{\rm min}=\sigma\frac{3}{\sqrt{2}}$, $\Delta {\cal E}_{\rm max}=\Delta {\cal E}_{\rm
hom}=\sigma\sqrt{6}$. The ratio $\frac{\Delta {\cal E}_{\rm max}}{\Delta {\cal E}_{\rm min}} = 2/\sqrt{3}=1.16$
and this narrow interval strictly limits the $\Delta {\cal E}$ variation. It has a simple graphical
interpretation. As is known, three real roots of a third order equation must fulfill the conditions (Cardan's
formulas) presented in Fig.1a, where the angle $\varphi$ is a function of the equation coefficients. The maximal
and minimal splittings $\Delta {\cal E}$ correspond to the solutions shown in Figs 1b and 1c, respectively.

Let us now pass to the Kramers ions with a half integer $J$. Here, two cases should be analyzed. Firstly, if an
even number of doublets $(2J+1)/2$ occurs the minimal $|J\rangle$ state splitting takes place when $(2J+1)/4$
doublets have the energy $\Delta {\cal E}_{\rm min}/2$, and the next $(2J+1)/4$ doublets the energy $-\Delta
{\cal E}_{\rm min}/2$. Then,
\begin{eqnarray*}
  \sigma^{2} &=& \frac{4[(2J+1)/4]\left(\Delta {\cal E}_{\rm min}/2\right)^2}{2J+1}, \qquad \qquad
  \Delta {\cal E}_{\rm min}=2\sigma.
\end{eqnarray*}
In turn, the maximal splitting, $\Delta {\cal E}_{\rm max}$, will appear if one of the doublets will be of
energy $\Delta {\cal E}_{\rm max}/2$, and the second one of energy $-\Delta {\cal E}_{\rm max}/2$ with all the
rest of levels with zero energy. This time
\begin{eqnarray*}
  \sigma^{2} &=& \frac{4\left(\Delta {\cal E}_{\rm max}/2\right)^2}{2J+1}, \qquad \qquad
  \Delta {\cal E}_{\rm max}=\sigma \sqrt{2J+1},
\end{eqnarray*}
and therefore now
\begin{eqnarray*}
 \frac{\Delta {\cal E}_{\rm max}}{\Delta{\cal E}_{\rm min}} &=& \frac{\sqrt{2J+1}}{2}.
\end{eqnarray*}

Secondly, for an odd number of Kramers doublets $\Delta {\cal E}_{\rm
min}=\frac{2\sigma(2J+1)}{\sqrt{(2J+3)(2J-1)}}\;$ and $\;\Delta {\cal E}_{\rm max}=\sigma\sqrt{2J+1}$, with
$\frac{\Delta {\cal E}_{\rm max}}{\Delta {\cal E}_{\rm min}} = \frac{1}{2}\sqrt{\frac{(2J+3)(2J-1)}{2J+1}}$.

The homogenous splitting $\Delta {\cal E}_{\rm hom}$ for an even number of doublets $(J=(4k+3)/2)$ and for an
odd number of doublets $(J=(4k+1)/2)$, where $k=0,1,\ldots$, amounts correspondingly to
$\sigma\sqrt{\frac{3(2J-1)}{J}}$ and $2\sigma\sqrt{\frac{3(2J-1)}{2J+3}}$. By way of example, if $J=9/2$ (i.e.
for five doublets), then $\Delta {\cal E}_{\rm min}=\sigma\frac{5}{\sqrt{6}}=2.04\sigma$, $\Delta {\cal E}_{\rm
max}=\sigma\sqrt{10}=3.16\sigma$, and $\Delta {\cal E}_{\rm hom}=2.83\sigma$. In the case of Kramers ions the
$\Delta {\cal E}$ variation range turns out to be smaller than that for non-Kramers ions, which is seen
comparing the $\Delta {\cal E}_{\rm min}$ and $\Delta {\cal E}_{\rm max}$ for $J=4$ and $J=9/2$. Finally, taking
the most extreme case of $J=15/2$ for $f$-electron configurations (e.g. for Dy$^{3+}$, Er$^{3+}$) with eight
doublets, we would obtain $\Delta {\cal E}_{\rm min}=\sigma\frac{16}{3\sqrt{7}}=2.02\sigma$, $\Delta {\cal
E}_{\rm max}=4\sigma$, and $\Delta {\cal E}_{\rm hom}=2.37\sigma$.

\section*{4. The new scale of the crystal-field strength. Comparison of both the scales
$S_{k}=\frac{1}{2k+1}M_{k}$ and $S_{k}^{\prime}=|{\cal H}_{\rm CF}^{(k)}|_{\rm av}$}

\subsection*{4.1. Average values of the axial parameter modules $|B_{k0}^{\prime}|_{\rm av}$ and
$|\sum_{k}B_{k0}^{\prime}|_{\rm av}$, where $k=2,4,6$ -- the rotational invariants of the equivalent ${\cal
H}_{\rm CF}$ parametrizations}
Rotating the reference frame by the two Euler angles $\alpha$ and $\beta$ we obtain all the equivalent ${\cal
H}_{\rm CF}$ parametrizations (with the accuracy to the third Euler angle $\gamma$ about the $z$ axis) [5,24].
Their axial parameters for a $2^{k}$-pole component are given as:
\begin{eqnarray}
  B_{k0}^{\prime} &=& \sum\limits_{q=-k}^{k}{\cal D}_{0q}^{(k)} (\alpha,\beta,0)B_{kq}=
  \sum\limits_{q=-k}^{k}C_{q}^{(k)}(\beta,\alpha)B_{kq} \nonumber\\
  &=& C_{0}^{(k)}(\beta)B_{k0} + 2\sum\limits_{q=1}^{k}C_{q}^{(k)}(\beta)\cos q(\alpha+\varphi_{q})
  |B_{kq}|,
\end{eqnarray}
where ${\cal D}_{0q}^{(k)} (\alpha,\beta,\gamma)$ are the middle row rotation matrix elements,
$C_{q}^{(k)}(\beta,\alpha)=\left(\frac{4\pi}{2k+1}\right)^{1/2}Y_{q}^{k}(\beta,\alpha)$ are the spherical
tensors, whereas $C_{q}^{(k)}(\beta)=(-1)^{q}\left[\frac{(k-q)!}{(k+q)!}\right]^{1/2}P_{k}^{q}(\cos\beta)$, and
$P_{k}^{q}(\cos\beta)$ are the associated Legendre functions, $B_{kq}=|B_{kq}|e^{{\rm i}q\varphi_{q}}$, and
$B_{k-q}=(-1)^{q}|B_{kq}|e^{-{\rm i}q\varphi_{q}}$. The primed parameters correspond to the transformed
parametrization while the unprimed to the initial one. It can be directly proved that
$\left(B_{k0}^{\prime}\right)_{\rm av}=0$ and $\left(\sum_{k}B_{k0}^{\prime}\right)_{\rm av}=0$, while the
average absolute values
\begin{eqnarray}
  \left|B_{k0}^{\prime}\right|_{\rm av}&=& \frac{1}{4\pi}\int\limits_{0}^{2\pi}\int\limits_{0}^{\pi}
  \left|B_{k0}^{\prime}|(\alpha,\beta)\right|\sin\beta d\beta d\alpha \;,\nonumber\\
  \left|\sum_{k}B_{k0}^{\prime}\right|_{\rm av}&=& \frac{1}{4\pi}\int\limits_{0}^{2\pi}\int\limits_{0}^{\pi}
  \left|\sum_{k}B_{k0}^{\prime}(\alpha,\beta)\right|\sin\beta d\beta d\alpha \;,
\end{eqnarray}
as the rotational group invariants are discriminants of the equivalent parametrizations classes [24]. By the
mean value we understand the magnitude averaged over all possible orientations of the reference frame, i.e. over
the solid angle $4\pi$. Interestingly, they can be used to estimate the CF strength independently of the
parametrization modulus.

\subsection*{4.2. Average values of the modules $\left|{\cal H}_{\rm CF}^{(k)}\right|_{\rm av}$ and
$\left|{\cal H}_{\rm CF}\right|_{\rm av}$}
Since the expression
\begin{eqnarray*}
  {\cal H}_{\rm CF}^{(k)} &=& \sum\limits_{q=-k}^{k}B_{kq}C_{q}^{(k)}(\beta,\alpha),
\end{eqnarray*}
where $\beta$ and $\alpha$ are the spherical angle coordinates in the central-ion reference system, is identical
with that for $B_{k0}^{\prime}$ (Eq.10), the following important identity holds
\begin{eqnarray}
  \left|{\cal H}_{\rm CF}^{(k)}\right|_{\rm av}&=& \frac{1}{4\pi}\int\limits_{0}^{2\pi}\int\limits_{0}^{\pi}
  \left|{\cal H}_{\rm CF}^{(k)}|(\alpha,\beta)\right|\sin\beta d\beta d\alpha=
  \left|B_{k0}^{\prime}\right|_{\rm av}.
\end{eqnarray}
The average value of the modulus of the $2^{k}$-pole ${\cal H}_{\rm CF}^{(k)}$ component turns out to be equal
to the average value of the modulus of the relevant axial parameter $B_{k0}$. This identity, Eq.12, obvious when
we properly interpret the rotation angles in both cases of averaging, associates
$\left|B_{k0}^{\prime}\right|_{\rm av}$ with the complementary measure of the CF strength $S_{k}^{\prime}$:
\begin{eqnarray}
  S_{k}^{\prime} &=& \left|{\cal H}_{\rm CF}^{(k)}\right|_{\rm av}= \left|B_{k0}^{\prime}\right|_{\rm av}
\end{eqnarray}
Although the expression for $S_{k}^{\prime}$ in the above form is limited to a given $2^{k}$-pole ${\cal H}_{\rm
CF}^{(k)}$ component, it may be generalized for the global ${\cal H}_{\rm CF}$
\begin{eqnarray}
  S^{\prime} &=& \left|\sum\limits_{k}{\cal H}_{\rm CF}^{(k)}\right|_{\rm av}=
  \left|\sum\limits_{k}B_{k0}^{\prime}\right|_{\rm av}.
\end{eqnarray}
Contrary to the conventional CF strengths $S_{k}$ and $S$ (Eqs 1,2), which are constant for all the iso-modular
parametrizations, the new strengths $S_{k}^{\prime}$ and $S^{\prime}$ calculated for the constant modulus (and
modules) change their magnitudes within certain ranges discussed in the next section. To compare both the
measures it is convenient to express $S_{k}^{\prime}$ in the product form $f_{k}\cdot M_{k}$, where $f_{k}$ is a
specified factor. Now, these two measures will be compatible if the factor $f_{k}$ is close to
$\sqrt{\frac{1}{2k+1}}$, Eq.1, i.e. to $0.447$, $0.333$ and $0.277$ for $k=2,4$ and $6$, respectively. This
compatibility is demonstrated in the next section, where a thorough discussion of the relation between both the
CF strength scales is provided, by way of example of the CF splitting of $p^{1}$, $d^{1}$ and $f^{1}$ electron
configurations with the spin-orbit coupling deliberately neglected , and the $^{3}H_{4}$ state for various
iso-modular ${\cal H}_{\rm CF}^{(k)}$s.

\subsection*{5. Computational results and discussion}
\subsection*{5.1. Crystal-field splitting of $p^{1}$, $d^{1}$ and $f^{1}$ electron configurations for various
iso-modular ${\cal H}_{\rm CF}^{(k)}$s, $k=2,4,6$.}

We consider the model results of interaction of any iso-modular ${\cal H}_{\rm CF}^{(k)}$ ($k=2,4,6$) with
$M_{k}=1$, on the initial states with well defined angular momentum quantum numbers. The magnitudes of all
quantities under discussion, i.e. the new CF strength parameters $S_{k}^{\prime}$, the total splittings $\Delta
E^{(k)}$, the second moments $\sigma_{k}^{2}$ of CF levels and the averages of the absolute values of the Stark
level energies $|E_{n}^{(k)}|_{\rm av}$ are given in $M_{k}$ units. Tables 1, 2 and 3 present a comprehensive
review of $S_{k}^{\prime}$ values for various iso-modular ${\cal H}_{\rm CF}^{(k)}$s, with $k=2,4,6$,
respectively. Correspondingly, these five, ten and eleven ${\cal H}_{\rm CF}^{(k)}$s compiled in Tables are the
representative ones including those with the highest and lowest $S_{k}^{\prime}$ values found during the survey.
No other ${\cal H}_{\rm CF}^{(k)}$s seem to yield $S_{k}^{\prime}$ out of these ranges. The strength parameters
$S_{k}^{\prime}$ change themselves within the rather narrow intervals: $0.368 - 0.385$, $0.251 - 0.287$ and
$0.195 - 0.239$, while the relevant $S_{k}$ are constant and equal to $0.447$, $0.333$ and $0.277$ for $k=2,4$
and $6$, respectively. The maximal $S_{k}^{\prime}$ parameters refer to the purely axial ${\cal H}_{\rm
CF}^{(k)}$s when $B_{k0}$s achieve 1. For other parametrizations this maximal value of 1 is not achieved in any
reference frame.

As implies from Tables 4, 5 and 6 there is a certain mapping between the above $S_{k}^{\prime}$ ranges and the
referring to them intervals of $\Delta E^{(k)}$ and $|E_{n}^{(k)}|_{\rm av}$. As is shown in the paper this
quantitative mapping is determined by the roots of the ${\cal H}_{\rm CF}^{(k)}$ matrix characteristic
polynomial, and the key part of the matrix elements is the product
$(-1)^{M_{J}}B_{kq}\left(\begin{array}{ccc}
                         J & k & J^{\prime} \\
                         -M_{J} & q & M_{J}^{\prime} \\
                         \end{array}\right)$.
The remaining factors coming into the matrix elements are common and play the role of a scaling factor. In the
below examples concerning the CF splitting of one-electron states with $J=J^{\prime}=l$ for $l=1,2$ and $3$, the
role of such a scaling factor play the double-bar matrix elements $\langle l||C^{(k)}||l \rangle$.

It should be pointed out, however, that the mappings $S_{k}^{\prime}\leftarrow\!\!\rightarrow \Delta E^{(k)}$,
$S_{k}^{\prime}\leftarrow\!\!\rightarrow |E_{n}^{(k)}|_{\rm av}$, $\Delta E^{(k)}\leftarrow\!\!\rightarrow
|E_{n}^{(k)}|_{\rm av}$ are neither straightforward nor explicit. With the increase of the initial state
degeneration $2J+1$ they become less clear due to a big variety of possible splitting schemes. Nevertheless, one
may presume a dominant tendency: the greater $S_{k}^{\prime}$ the greater $|E_{n}^{(k)}|_{\rm av}$ and the
lesser $\Delta E^{(k)}$ (Type I splittings). In the reverse case, i.e. for a small $S_{k}^{\prime}$, Type II
splittings are expected. However, such reasoning does not take into account the unique characteristics of the
Hamiltonian averages $\left|{\cal H}_{\rm CF}^{(k)}\right|_{\rm av}$, and the space density distribution of
unpaired electrons in the states of various $J$. From this point of view the analysis of Tables 4, 5 and 6 seems
to be instructive, indeed.

On the other hand, the allowed spans of the $\Delta {\cal E}^{(k)}$ for a fix $M_{k}$, i.e. $\sigma_{k}$ are
known. In the light of the above mapping it turns out that not all of these values $\Delta {\cal E}^{(k)}$, and
corresponding to them splitting schemes, can actually occur. Namely, depending on the initial state quantum
number $J$ and the multipole's rank $k$ some specified limitations of the $\Delta E^{(k)}$ are observed (Tables
4, 5 and 6). They are listed briefly below.

For $l=1$ ($p$-electron) and $k=2$ the full nominal range of the $\Delta {\cal E}^{(2)}$ and all splittings of
Types I, II and III are admitted. More particularly, $\Delta E^{(2)}$ can vary from $0.600\;M_{2}$ to
$0.693\;M_{2}$ (Table 4).

For $l=2$ ($d$-electron) and $k=2$ the magnitude of $\Delta E^{(2)}$ is constant and equals $0.572\;M_{2}$ for
each iso-modular ${\cal H}_{\rm CF}^{(2)}$ what corresponds to splittings similar to those of Type I. Other
splittings, including e.g. $\Delta {\cal E}^{(2)}_{\rm hom}$ are impossible in this case (Table 5).

For $l=2$ ($p$-electron) and $k=4$ again the full nominal range of the $\Delta {\cal E}^{(4)}$ is allowed
beginning from the smallest $0.363M_{4}$ for the cubic ${\cal H}_{\rm CF}^{(4)}$, up to the biggest $0.564
M_{4}$ for ${\cal H}_{\rm CF}^{(4)}=\frac{1}{\sqrt{2}}B_{44}C_{4}^{(4)}+\frac{1}{\sqrt{2}}B_{4-4}C_{-4}^{(4)}$
(Table 5).

For $l=3$ ($f$-electron) and the value of $k=2$ $\Delta E^{(2)}$ weakly depends on $S_{2}^{\prime}$, varying in
all its range merely from $0.600M_{2}$ to $0.608M_{2}$, i.e. somewhat below the $\Delta {\cal E}^{(2)}_{\rm
hom}$ (Type III splittings) (Table 6).

Next, for $l=3$ and $k=4$, the possible $\Delta E^{(4)}$ varies within the range from $0.358M_{4}$ to
$0.482M_{4}$, i.e. around the $\Delta {\cal E}^{(4)}_{\rm hom}$ (Table 6).

Finally, for $l=3$ and $k=6$ the allowed $\Delta E^{(6)}$ varies from $0.326M_{6}$ to $0.501M_{6}$ covering the
majority of the nominal range together with its upper limit, but excluding the smallest splittings (Table 6).

The obtained results may be generalized for states with $J$ or $L$ equal to 1, 2 or 3, multiplying $\Delta
E^{(k)}$ and $|E_{n}^{(k)}|_{\rm av}$ by the scaling factors $\langle J||C^{(k)}||J\rangle$ or $\langle
L||C^{(k)}||L\rangle$.

\subsection*{5.2. Crystal-field splitting of $^{3}H_{4}$ state in various iso-modular
${\cal H}_{\rm CF}^{(k)}$s, $k=2,4,6$.}
Let us end up with the analysis of splitting of nine-fold degenerate $^{3}H_{4}$ state subjected to the
iso-modular ${\cal H}_{\rm CF}^{(k)}$s enclosed in Tables 1, 2 and 3. Table 7 shows the correlation between
$S_{k}^{\prime}$, $\Delta E^{(k)}$ and $|E_{n}^{(k)}|_{\rm av}$. The scaling factors $\langle
J\!=\!4||C^{(k)}||J\!=\!4\rangle=\langle J\!=\!4||U^{(k)}||J\!=\!4\rangle\langle f||C^{(k)}||f\rangle$, required
here due to the coupled initial state $(L=5, S=1, J=4)$, are equal to $-1.2365$, $-0.7389$ and $0.7706$ for
$k=2,4,6$, respectively. Hence $\sigma_{2}=0.184M_{2}$, $\sigma_{4}=0.082M_{4}$ and $\sigma_{6}=0.071M_{6}$,
while the global second moment of the Stark levels takes the form
\begin{eqnarray*}
  \sigma^{2}&=& \frac{1}{9}\left[\frac{1}{5}(-1.2365)^{2}M_{2}^{2}+\frac{1}{9}(-0.7389)^{2}M_{4}^{2}
  +\frac{1}{13}(0.7706)^{2}M_{6}^{2}\right].
\end{eqnarray*}
The ranges of the formally allowed $\Delta {\cal E}^{(k)}$ corresponding to the above second moments
$\sigma_{2},\sigma_{4},\sigma_{6}$ are marked in Fig.2 by the solid lines.

We can see in Fig.2 that from the set of all potentially allowed total splittings $\Delta {\cal E}^{(k)}$ only
certain $\Delta E^{(k)}$ may be realized (those between the dashed lines), and consequently, only certain
splitting schemes (roughly between Types I and III) may occur. For instance, in the case of all the three
effective multipoles neither $\Delta {\cal E}^{(k)}_{\rm max}$ nor $\Delta {\cal E}^{(k)}_{\rm min}$ are
possible, while $\Delta {\cal E}^{(k)}_{\rm hom}$ can appear solely in the case of $2^{6}$-pole. Based on Table
7 it is seen also that for all the three effective ${\cal H}_{\rm CF}^{(k)}$s the biggest $\Delta E^{(k)}$ are
achieved for intermediate $S_{k}^{\prime}$ values.
\subsection*{6. Conclusions}

The conventional scales $S_{k}$ or $S$ with the associated second moments of the CF levels, $\sigma_{k}$ or
$\sigma$, do not distinguish the iso-modular ${\cal H}_{\rm CF}^{(k)}$ or ${\cal H}_{\rm CF}$ parametrizations,
which, however, can be differentiated by an another scale -- the spherically averaged
$S_{k}^{\prime}=\left|{\cal H}_{\rm CF}^{(k)}\right|_{\rm av}$ and $S^{\prime}=\left|{\cal H}_{\rm
CF}\right|_{\rm av}$. It is proved that the $S_{k}^{\prime}$ variation ranges for all the iso-modular
parametrizations are limited and lie slightly below the relevant $S_{k}$ magnitudes. The span of these ranges
amounts to 5, 10 and 20\% of their values for $k=2,4$ and $6$, respectively. There exists a direct mapping of
$S_{k}^{\prime}$ ranges into the total splitting $\Delta E^{(k)}$ ranges and $|E_{n}^{(k)}|_{\rm av}$ intervals,
which may be interpreted more clearly for the initial states with low degeneration. Such mapping allows to
estimate the total splittings $\Delta E^{(k)}$ or $\Delta E$ to be expected and characterize their spectrum. It
is shown that not all the nominally admitted total $\Delta {\cal E}^{(k)}$ or $\Delta {\cal E}$ splittings
determined by the modules $M_{k}$ or $M$, i.e. the second moments $\sigma_{k}$ or $\sigma$, can actually occur.
This essentially confines the set of the allowed splitting schemes.

\newpage

\renewcommand{\baselinestretch}{1}


\newpage
\renewcommand{\baselinestretch}{1}

\begin{table}[htb]
\begin{center}
\caption{The spherical averages of five representative iso-modular ${\cal H}_{\rm CF}^{(2)}$s,
$\;S_{2}^{\prime}=|{\cal H}_{\rm CF}^{(2)}|_{av}$, acc. to Eqs 11-13, expressed in $M_{2}$ units. Only $B_{2q}$
CFPs are given, $B_{2-q}=(-1)^{q}B_{2q}^{\ast}$} \label{tab} \vspace*{0.7cm}
\begin{tabular}{lcccccc}
\hline
No.& &\multicolumn{3}{c}{${\cal H}_{\rm CF}^{(2)}$ composition}& &$S_{2}^{\prime}=|{\cal H}_{\rm CF}^{(2)}|_{\rm av}$\\
\cline{3-5}
& &$B_{20}$&$B_{21}$&$B_{22}$& & \\
\hline
1& &1&0&0& &0.385\\
2& &$\frac{1}{\sqrt{5}}$&$\frac{1}{\sqrt{5}}$&$-\frac{1}{\sqrt{5}}$& &0.381\\
3& &$\frac{1}{\sqrt{5}}$&$\frac{1}{\sqrt{5}}$&$-\frac{1}{\sqrt{5}}$& &0.374\\
4& &$\frac{1}{\sqrt{5}}$&$\frac{1}{\sqrt{5}}\;\;{\rm e}^{{\rm i}\pi/4}$&$\frac{1}{\sqrt{5}}$& &0.369\\
5& &0&0&$\frac{1}{\sqrt{2}}$& &0.368\\
\hline
\end{tabular}
\end{center}
\end{table}

\newpage
\renewcommand{\baselinestretch}{1}

\begin{table}[htb]
\begin{center}
\caption{The spherical averages of ten representative iso-modular ${\cal H}_{\rm CF}^{(4)}$s,
$\;S_{4}^{\prime}=|{\cal H}_{\rm CF}^{(4)}|_{av}$, acc. to Eqs 11-13, expressed in $M_{4}$ units. Only $B_{4q}$
CFPs are given, $B_{4-q}=(-1)^{q}B_{4q}^{\ast}$} \label{tab} \vspace*{0.7cm}
\begin{tabular}{lcccccccc}
\hline
No.& &\multicolumn{5}{c}{${\cal H}_{\rm CF}^{(4)}$ composition}& &$S_{4}^{\prime}=|{\cal H}_{\rm CF}^{(4)}|_{\rm av}$\\
\cline{3-7}
& &$B_{40}$&$B_{41}$&$B_{42}$&$B_{43}$&$B_{44}$&  & \\
\hline
1& &1&0&0&0&0& &0.287\\
2& &$\frac{1}{2}\sqrt{\frac{7}{3}}$&0&0&0&$\frac{1}{2}\sqrt{\frac{5}{6}}$& &0.280\\
3& &$\frac{1}{3}$&$\frac{1}{3}$&$\frac{1}{3}\;{\rm e}^{{\rm i}\pi/2}$&$\frac{1}{3}$&$\frac{1}{3}$& &0.277\\
4& &$\frac{1}{3}$&$-\frac{1}{3}$&$\frac{1}{3}$&$\frac{1}{3}$&$\frac{1}{3}$& &0.276\\
5& &$\frac{1}{3}$&$\frac{1}{3}$&$\frac{1}{3}$&$-\frac{1}{3}\;{\rm e}^{{\rm i}\pi/2}$&$\frac{1}{3}$& &0.273\\
6& &0&0&$\frac{1}{\sqrt{2}}$&0&0& &0.269\\
7& &0&0&0&$\frac{1}{\sqrt{2}}$&0& &0.266\\
8& &$\frac{1}{3}$&$\frac{1}{3}$&$\frac{1}{3}$&$\frac{1}{3}$&$\frac{1}{3}$& &0.265\\
9& &$\frac{1}{3}$&$\frac{1}{3}\;{\rm e}^{{\rm i}\pi/4}$&$\frac{1}{3}$&$\frac{1}{3}$&$\frac{1}{3}$& &0.261\\
10& &0&0&0&0&$\frac{1}{\sqrt{2}}$ &&0.251\\
\hline
\end{tabular}
\end{center}
\end{table}

\newpage
\renewcommand{\baselinestretch}{1}

\begin{table}[htb]
\begin{center}
\caption{The spherical averages of eleven representative iso-modular ${\cal H}_{\rm CF}^{(6)}$s,
$\;S_{6}^{\prime}=|{\cal H}_{\rm CF}^{(6)}|_{av}$, acc. to Eqs 11-13, expressed in $M_{6}$ units. Only $B_{6q}$
CFPs are given, $B_{6-q}=(-1)^{q}B_{6q}^{\ast}$} \label{tab} \vspace*{0.7cm}
\begin{tabular}{lcccccccccc}
\hline
No.& &\multicolumn{7}{c}{${\cal H}_{\rm CF}^{(6)}$ composition}& &$S_{6}^{\prime}=|{\cal H}_{\rm CF}^{(6)}|_{\rm av}$\\
\cline{3-9}
& &$B_{60}$&$B_{61}$&$B_{62}$&$B_{63}$&$B_{64}$&$B_{65}$&$B_{66}$&  & \\
\hline
1& &1&0&0&0&0&0&0& &0.239\\
2& &$\frac{1}{\sqrt{13}}$&$\frac{1}{\sqrt{13}}$&$\frac{1}{\sqrt{13}}$&$\frac{1}{\sqrt{13}}$
&$-\frac{1}{\sqrt{13}}$&$\frac{1}{\sqrt{13}}$&$\frac{1}{\sqrt{13}}$& &0.231\\
3& &$\frac{1}{\sqrt{13}}$&$\frac{1}{\sqrt{13}}$&$-\frac{1}{\sqrt{13}}$&$\frac{1}{\sqrt{13}}$
&$\frac{1}{\sqrt{13}}$&$\frac{1}{\sqrt{13}}$&$\frac{1}{\sqrt{13}}$& &0.228\\
4& &$\frac{1}{\sqrt{13}}$&$\frac{1}{\sqrt{13}}$&$-\frac{1}{\sqrt{13}}$&$\frac{1}{\sqrt{13}}$
&$-\frac{1}{\sqrt{13}}$&$\frac{1}{\sqrt{13}}$&$\frac{1}{\sqrt{13}}$& &0.227\\
5& &$-\frac{1}{\sqrt{13}}$&$\frac{1}{\sqrt{13}}$&$\frac{1}{\sqrt{13}}$&$\frac{1}{\sqrt{13}}$
&$\frac{1}{\sqrt{13}}$&$\frac{1}{\sqrt{13}}$&$\frac{1}{\sqrt{13}}$& &0.225\\
6& &$\frac{1}{\sqrt{13}}$&$-\frac{1}{\sqrt{13}}$&$\frac{1}{\sqrt{13}}$&$\frac{1}{\sqrt{13}}$
&$\frac{1}{\sqrt{13}}$&$\frac{1}{\sqrt{13}}$&$\frac{1}{\sqrt{13}}$& &0.223\\
7& &$\frac{1}{2\sqrt{2}}$&0&0&0&$\pm\frac{\sqrt{7}}{4}$&0&0& &0.223\\
8& &$\frac{1}{\sqrt{13}}$&$\frac{1}{\sqrt{13}}$&$\frac{1}{\sqrt{13}}$&$\frac{1}{\sqrt{13}}$
&$\frac{1}{\sqrt{13}}$&$\frac{1}{\sqrt{13}}$&$-\frac{1}{\sqrt{13}}$& &0.222\\
9& &0&$\frac{1}{\sqrt{2}}$&0&0&0&0&0& &0.219\\
10& &$\frac{1}{\sqrt{13}}$&$\frac{1}{\sqrt{13}}$&$\frac{1}{\sqrt{13}}$&$\frac{1}{\sqrt{13}}$
&$\frac{1}{\sqrt{13}}$&$\frac{1}{\sqrt{13}}$&$\frac{1}{\sqrt{13}}$& &0.213\\
11& &0&0&0&0&0&0&$\frac{1}{\sqrt{2}}$& &0.195\\
\hline
\end{tabular}
\end{center}
\end{table}

\newpage
\renewcommand{\baselinestretch}{1}

\begin{table}[htb]
\begin{center}
\caption{The total crystal-field splitting $\Delta E^{(2)}$ of $p^{1}$ configuration (with $ls$ coupling
neglected) and the average absolute values of $E_{n}^{(2)}$ in the crystal-field potentials given in Table 1.
All the values are given in $M_{2}$ units.} \label{tab} \vspace*{0.7cm}
\begin{tabular}{lcccc}
\hline
No.& &$\left|{\cal H}_{\rm CF}^{(2)}\right|_{\rm av}$&$\Delta E^{(2)}$& $\left|E_{n}^{(2)}\right|_{\rm av}$\\
\hline
1& &0.385& 0.600 &0.267\\
2& &0.381& 0.656 &0.262\\
3& &0.374& 0.683 &0.250\\
4& &0.369& 0.692 &0.239\\
5& &0.368& 0.693 &0.231\\
\hline \multicolumn{4}{c}{{\small $\sigma_{2}=\sqrt{2/25}=0.283$, $\Delta {\cal E}^{(2)}_{\rm min}=0.600$,
$\Delta {\cal E}^{(2)}_{\rm hom}=\Delta {\cal E}^{(2)}_{\rm max}=0.693$ (Eqs 6-9)}}
\end{tabular}
\end{center}
\end{table}

\newpage
\renewcommand{\baselinestretch}{1}

\begin{table}[htb]
\begin{center}
\caption{The total crystal-field splittings $\Delta E^{(2)}$ and $\Delta E^{(4)}$ of $d^{1}$ configuration (with
$ls$ coupling neglected) and the average absolute values of $E_{n}^{(2)}$ and $E_{n}^{(4)}$ in the crystal-field
potentials given in Tables 1 and 2. All the values are given in $M_{2}$ and $M_{4}$ units, respectively.}
\label{tab} \vspace*{0.7cm}
\begin{tabular}{lcccc}
\hline
No.& &$\left|{\cal H}_{\rm CF}^{(2)}\right|_{\rm av}$&$\Delta E^{(2)}$& $\left|E_{n}^{(2)}\right|_{\rm av}$\\
\hline
1& &0.385& 0.572 &0.229\\
2& &0.381& 0.572 &0.227\\
3& &0.374& 0.572 &0.221\\
4& &0.369& 0.572 &0.217\\
5& &0.368& 0.572 &0.213\\
\hline \multicolumn{4}{c}{{\small $\sigma_{2}=\sqrt{2/35}=0.239$, $\Delta {\cal E}^{(2)}_{\rm min}=0.488$,
$\Delta
{\cal E}^{(2)}_{\rm hom}=0.676$, $\Delta {\cal E}^{(2)}_{\rm max}=0.756$ (Eqs 6-9)}}\\
\\
\hline
 & &$\left|{\cal H}_{\rm CF}^{(4)}\right|_{\rm av}$&$\Delta E^{(4)}$& $\left|E_{n}^{(4)}\right|_{\rm av}$\\
\hline
6& &0.287& 0.476 &0.152\\
7& &0.280& 0.363 &0.174\\
8& &0.277& 0.449 &0.169\\
9& &0.276& 0.437 &0.169\\
10& &0.273& 0.463 &0.164\\
11& &0.269& 0.426 &0.159\\
12& &0.266& 0.398 &0.159\\
13& &0.265& 0.549 &0.137\\
14& &0.261& 0.555 &0.132\\
15& &0.251& 0.564 &0.113\\
\hline \multicolumn{4}{c}{{\small $\sigma_{4}=\sqrt{2/63}=0.178$, $\Delta {\cal E}^{(4)}_{\rm min}=0.363$,
$\Delta {\cal E}^{(4)}_{\rm hom}=0.503$, $\Delta {\cal E}^{(4)}_{\rm max}=0.564$ (Eqs 6-9)}}
\end{tabular}
\end{center}
\end{table}

\newpage
\renewcommand{\baselinestretch}{0.8}
\begin{table}[htb]
\begin{center}
\caption{The total crystal-field splittings $\Delta E^{(2)}$, $\Delta E^{(4)}$ and $\Delta E^{(6)}$ of $f^{1}$
configuration (with $ls$ coupling neglected) and the average absolute values of $E_{n}^{(2)}$, $E_{n}^{(4)}$ and
$E_{n}^{(6)}$ in the crystal-field potentials given in Tables 1, 2 and 3. All the values are given in $M_{2}$,
$M_{4}$ and $M_{6}$ units, respectively.} \label{tab} \vspace*{0.7cm}
\begin{tabular}{lcccc}
\hline
No.& &$\left|{\cal H}_{\rm CF}^{(2)}\right|_{\rm av}$&$\Delta E^{(2)}$& $\left|E_{n}^{(2)}\right|_{\rm av}$\\
\hline
1& &0.385& 0.600 &0.190\\
2& &0.381& 0.603 &0.191\\
3& &0.374& 0.607 &0.192\\
4& &0.369& 0.608 &0.193\\
5& &0.368& 0.608 &0.193\\
\hline \multicolumn{4}{c}{{\small $\sigma_{2}=\sqrt{4/75}=0.231$, $\Delta {\cal E}^{(2)}_{\rm min}=0.467$,
$\Delta
{\cal E}^{(2)}_{\rm hom}=0.693$, $\Delta {\cal E}^{(2)}_{\rm max}=0.864$ (Eqs 6-9)}}\\
\\
\hline
 & &$\left|{\cal H}_{\rm CF}^{(4)}\right|_{\rm av}$&$\Delta E^{(4)}$& $\left|E_{n}^{(4)}\right|_{\rm av}$\\
\hline
6& &0.287& 0.394 &0.121\\
7& &0.280& 0.417 &0.119\\
8& &0.277& 0.449 &0.116\\
9& &0.276& 0.458 &0.117\\
10& &0.273& 0.464 &0.115\\
11& &0.269& 0.478 &0.113\\
12& &0.266& 0.482 &0.115\\
13& &0.265& 0.399 &0.129\\
14& &0.261& 0.363 &0.129\\
15& &0.251& 0.358 &0.131\\
\hline \multicolumn{4}{c}{{\small $\sigma_{4}=\sqrt{2/99}=0.142$, $\Delta {\cal E}^{(4)}_{\rm min}=0.287$,
$\Delta
{\cal E}^{(4)}_{\rm hom}=0.426$, $\Delta {\cal E}^{(4)}_{\rm max}=0.531$ (Eqs 6-9)}}\\
\\
\hline
 & &$\left|{\cal H}_{\rm CF}^{(6)}\right|_{\rm av}$&$\Delta E^{(6)}$& $\left|E_{n}^{(6)}\right|_{\rm av}$\\
\hline
16& &0.239& 0.408 &0.107\\
17& &0.231& 0.326 &0.130\\
18& &0.228& 0.365 &0.120\\
19& &0.227& 0.379 &0.123\\
20& &0.225& 0.445 &0.106\\
21& &0.223& 0.420 &0.112\\
22& &0.223& 0.346 &0.127\\
23& &0.222& 0.422 &0.112\\
24& &0.219& 0.468 &0.097\\
25& &0.213& 0.481 &0.097\\
26& &0.195& 0.501 &0.072\\
\hline \multicolumn{4}{c}{{\small $\sigma_{6}=\frac{10}{13\sqrt{33}}=0.134$, $\Delta {\cal E}^{(6)}_{\rm
min}=0.271$, $\Delta {\cal E}^{(6)}_{\rm hom}=0.402$, $\Delta {\cal E}^{(6)}_{\rm max}=0.501$ (Eqs 6-9)}}
\end{tabular}
\end{center}
\end{table}


\newpage

\renewcommand{\baselinestretch}{0.8}
\begin{table}[htb]
\begin{center}
\caption{The total crystal-field splitting $\Delta E^{(2)}$, $\Delta E^{(4)}$ and $\Delta E^{(6)}$ of the
$^{3}H_{4}$ state and the average absolute values of $E_{n}^{(2)}$, $E_{n}^{(4)}$ and $E_{n}^{(6)}$ in the
crystal-field potentials given in Tables 1, 2 and 3. All the values are given in $M_{2}$, $M_{4}$ and $M_{6}$
units, respectively.} \label{tab} \vspace*{0.7cm}
\begin{tabular}{lcccc}
\hline
No.& &$\left|{\cal H}_{\rm CF}^{(2)}\right|_{\rm av}$&$\Delta E^{(2)}$& $\left|E_{n}^{(2)}\right|_{\rm av}$\\
\hline
1& &0.385& 0.504 &0.163\\
2& &0.381& 0.543 &0.170\\
3& &0.374& 0.562 &0.164\\
4& &0.369& 0.560 &0.164\\
5& &0.368& 0.524 &0.155\\
\hline \multicolumn{4}{c}{{\small $\sigma_{2}=0.184$, $\Delta {\cal E}^{(2)}_{\rm min}=0.370$, $\Delta {\cal
E}^{(2)}_{\rm hom}=0.570$, $\Delta {\cal E}^{(2)}_{\rm max}=0.781$ (Eqs 6-9)}}
\\
\\
\hline
 & &$\left|{\cal H}_{\rm CF}^{(4)}\right|_{\rm av}$&$\Delta E^{(4)}$& $\left|E_{n}^{(4)}\right|_{\rm av}$\\
\hline
6& &0.287& 0.215 &0.079\\
7& &0.280& 0.227 &0.073\\
8& &0.277& 0.249 &0.070\\
9& &0.276& 0.241 &0.071\\
10& &0.273& 0.232 &0.073\\
11& &0.269& 0.230 &0.064\\
12& &0.266& 0.232 &0.067\\
13& &0.265& 0.231 &0.077\\
14& &0.261& 0.229 &0.076\\
15& &0.251& 0.196 &0.077\\
\hline \multicolumn{4}{c}{{\small $\sigma_{4}=0.082$, $\Delta {\cal E}^{(4)}_{\rm min}=0.165$, $\Delta {\cal
E}^{(4)}_{\rm hom}=0.254$, $\Delta {\cal E}^{(4)}_{\rm max}=0.348$ (Eqs 6-9)}}
\\
\\
\hline
 & &$\left|{\cal H}_{\rm CF}^{(6)}\right|_{\rm av}$&$\Delta E^{(6)}$& $\left|E_{n}^{(6)}\right|_{\rm av}$\\
\hline
16& &0.239& 0.202 &0.058\\
17& &0.231& 0.192 &0.069\\
18& &0.228& 0.249 &0.051\\
19& &0.227& 0.212 &0.060\\
20& &0.225& 0.224 &0.062\\
21& &0.223& 0.233 &0.059\\
22& &0.223& 0.245 &0.053\\
23& &0.222& 0.233 &0.061\\
24& &0.219& 0.258 &0.052\\
25& &0.213& 0.208 &0.059\\
26& &0.195& 0.206 &0.058\\
\hline \multicolumn{4}{c}{{\small $\sigma_{6}=0.071$, $\Delta {\cal E}^{(6)}_{\rm min}=0.143$, $\Delta {\cal
E}^{(6)}_{\rm hom}=0.220$, $\Delta {\cal E}^{(6)}_{\rm max}=0.301$ (Eqs 6-9)}}
\end{tabular}
\end{center}
\end{table}


\newpage
\noindent FIGURE CAPTIONS:

\noindent Fig.1. Crystal-field splitting of $|J\!=\!1\rangle$ state -- geometrical interpretation ($x$ is the
energy) a) general case: $\Delta {\cal E}^{(2)}=x_{1}-x_{2}$, b) $\Delta {\cal E}^{(2)}_{\rm min}=x_{1}-x_{2}$,
and c) $\Delta {\cal E}^{(2)}_{\rm hom}=\Delta {\cal E}^{(2)}_{\rm max}=x_{2}-x_{3}$.

\noindent Fig.2. Nominally allowed $\Delta {\cal E}^{(k)}$ (bold solid borders) and the actual $\Delta E^{(k)}$
(dashed borders) ranges of the total splittings of the $^{3}H_{4}$ state subjected to the iso-modular ${\cal
H}_{CF}^{(k)}$. The $\Delta {\cal E}^{(k)}_{\rm hom}$ are also given (thin solid lines).


\newpage

\begin{center}
\vspace*{5cm}
\includegraphics[width=17cm]{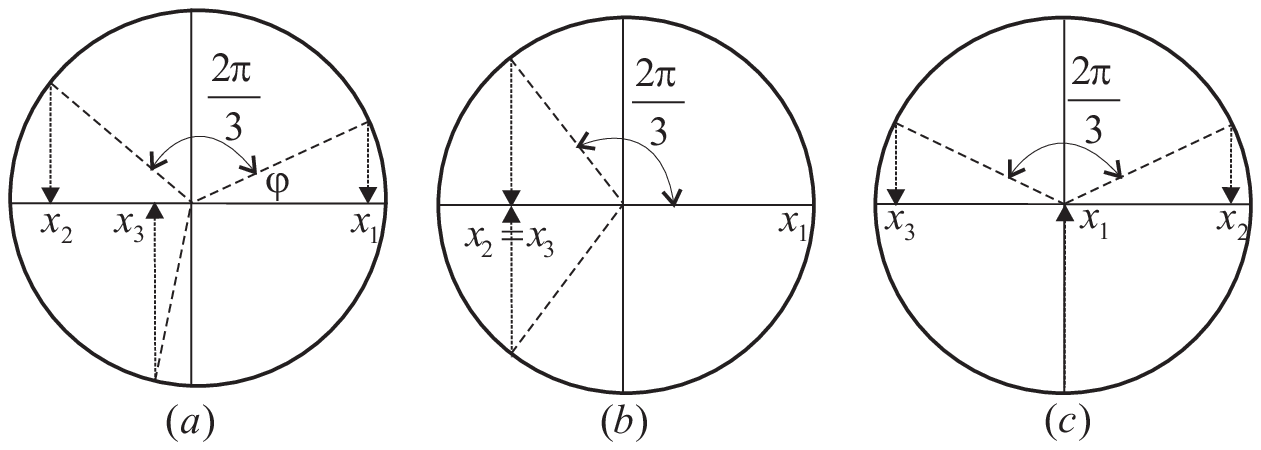}

\vspace*{2cm} \noindent Fig.1

\newpage
\vspace*{5cm}
\includegraphics[width=15cm]{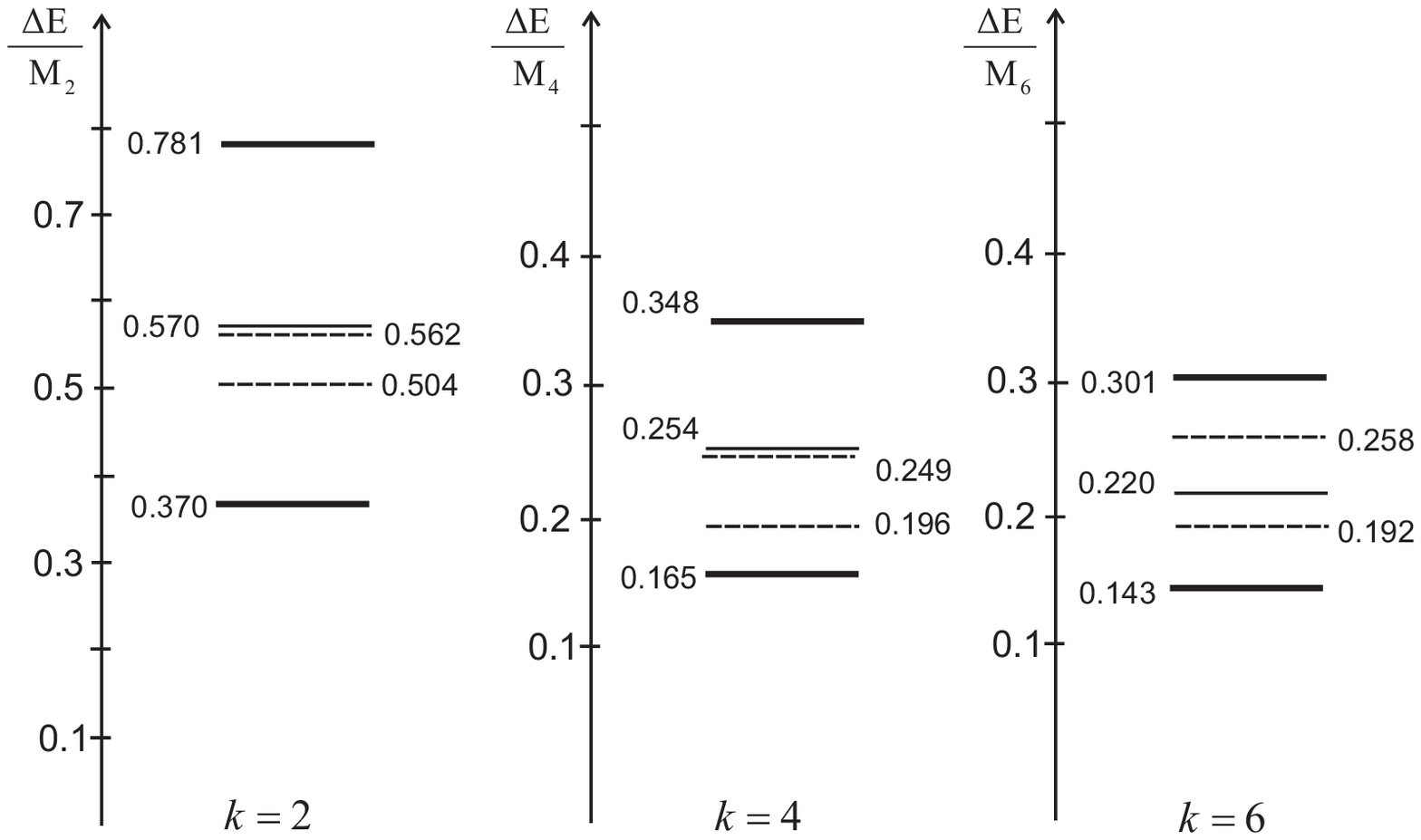}

\vspace*{2cm} \noindent Fig.2

\end{center}

\end{document}